\documentclass[aps,prc,twocolumn,superscriptaddress,showpacs,nofootinbib,floatfix]{revtex4}

\usepackage{graphicx} \usepackage{dcolumn} \usepackage{bm}
\newcommand{\beq}{\begin{equation}}\newcommand{\eeq}[1]{\label{#1}
\end{equation}}\newcommand{\beqar}{\begin{eqnarray}}\newcommand{\eeqar}[1]
{\label{#1}
\end{eqnarray}}\newcommand{\bmath}{\begin{displaymath}}\newcommand{\emath}{\end{displaymath}}\newcommand{\bitem}{\begin{itemize}}\newcommand{\eitem}{\end{itemize}}

\begin{document}

\title{Model for hypernucleus production in heavy ion collisions}

\author{V. Topor Pop and S. Das Gupta}

\affiliation{Physics Department, McGill University, 
Montr{\'e}al, Canada H3A 2T8}

\date{\today}

\begin{abstract}

We estimate the production cross sections of hypernuclei in projectile
like fragment (PLF) in heavy ion collisions. 
The discussed scenario for the formation cross section of 
$\Lambda$ hypernucleus is: (a) $\Lambda$ 
particles are produced in the participant region but have a considerable
rapidity spread and (b) $\Lambda$ with rapidity close to that of the
PLF and total momentum (in the rest system of PLF) up to Fermi motion  
can then be trapped and produce hypernuclei.  
The process (a) is 
considered here within Heavy Ion Jet Interacting Generator
({\small HIJING/B\=B}) model 
and the process (b) in the canonical thermodynamic model (CTM).
We estimate the production cross sections for a 
hypernucleus $^{A_{F}}_{\Lambda}Z$ where $Z$ = 1, 2, 3 and 4 for C + C at 
total nucleon-nucleon center of mass (c.m.) energy $\sqrt{s_{NN}}$ = 3.7
 GeV, and for Ne+Ne and Ar+Ar collisions at $\sqrt{s_{NN}}$ = 5.0 GeV.
By taking into account explicitly the impact parameter dependence of
the colliding systems, it is found that the cross section is 
different from that predicted by the coalescence model 
and large discrepancy
is obtained for $^6_{\Lambda}$He and $^9_{\Lambda}$Be hypernuclei. 

\end{abstract}


\maketitle

\section{Introduction}

Hypernuclear production in reactions between heavy nuclei was 
first theoretically proposed by Kerman and Weiss \cite{Weiss73}.
They found that relativistic 
heavy ion collisions (rHIC) offer the best possibility to create
exotic finite nuclear system with finite strangeness.
However, the experimental results have so far been rather limited
due to short lifetime of hypernuclei 
\cite{avramenko92,alberico02,averyanov08}, 
which impedes their detection.
Since the mechanism of the heavy ion induced hypernuclear reaction 
is not well understood, this field of research attracted mainly 
theoretical interest, see, e.g., Refs. 
\cite{Wakai88,Asai84,Bando89,Ko85,Bando90,Bando2,Sano94,Sato81,Sano88,Dover94,Rudy95,Greiner95,Carsten98,Mosel06,Botvina07,Subal09,Gaitanos09}.

The future experimental projects as planned  
at the Facility for Anti-protons and Ion Research (GSI-FAIR, Germany)
\cite{HypHI05,HypHI06,Saito07,fukuda07,Gianotti07,Friese07} 
and the Nuclotron-based Ion Collider Facility (NICA), 
at Joint Institute for Nuclear Research (JINR),
Dubna, Russia \cite{Toneev07},
aim to look for light proton/neutron rich exotic 
hypernuclei, and to extend their study to
heavier hypernuclei (which can be produced only in rHIC), toward
protons and neutron drip lines.
In addition, there are special light hypernuclei of interest, whose properties
are dictated by their unusual structure \cite{sakaguchi09}, 
e.g., nuclei with the hyperon
halo ($^3_{\Lambda}$H); neutron-rich hypernuclei; nuclei with
an unstable core, where the hyperon is a sort of ``{\em glue}''
ensuring stability ($^6_{\Lambda}$H, $^6_{\Lambda}$He, $^8_{\Lambda}$He). 
Recently at RHIC BNL (Relativistic Heavy Ion Collider at Brookhaven National
Laboratory) the observation of (anti)hypertritons has been reported
in nuclear collisions \cite{star2010},\cite{Xu2010}.

The research program at the SPS CERN-Geneva 
(Super Proton Synchroton at
European Organization for Nuclear Research) and at 
RHIC BNL will cover the energy range $\sqrt{s_{NN}}$ = 5-50 GeV
from below up to well above the energy of the onset 
of deconfinement (expected approximately at $\sqrt{s_{NN}}$ = 6-8 GeV). 
Moreover, HypHI-collaboration \cite{HypHI05,HypHI06} and 
the project Multi-Purpose Detector (MPD/NICA) \cite{Toneev07}  
will cover the 
relevant energies $\sqrt{s_{NN}}$ = 2.7-9.4 GeV.
These experiments aim to measure the production of hypernuclei 
in energetic collisions between light nuclei, since
their identification via the weak decay of the hyperon
into pion is much easier for such systems.

In this paper a hybrid model based on participant-spectator picture 
and combined with canonical thermodynamic (CTM) model is used 
to determine, in high energy heavy-ion collisions, the probability
of forming a hypernucleus and to estimate its production cross section.
For rHIC in 3-10 GeV energy range the following scenario
(supported by experiments) applies.  For a general impact parameter
there is a region of violent collision called the participating region.
In addition, there is a mildly excited projectile like fragment (PLF)
and also a mildly excited target like fragment (TLF). Physics of both
PLF and TLF are similar for symmetric collisions; here we concentrate 
our analysis on PLF. 
Because of excitation energy (usually characterized by
a temperature, $T$) PLF will break up into many fragments 
and the velocities
of these fragments are centered around the velocity of the projectile.
In fixed target experiments they are emitted in a forward cone and
are more easily recognizable. In the participating
region, apart from original neutrons and protons, others 
particles (pions, kaons, $\Lambda$'s, etc.) are produced. 
The produced $\Lambda$ particles have an extended rapidity
range.  
 Those produced in the rapidity range close to that of the
projectile and having total momenta in the PLF frame 
up to Fermi momentum
($p_{\rm tot} < 250$ MeV/c) can be trapped in the PLF 
and form hypernuclei.  
The present problem has also been analyzed previously 
(see Refs. \cite{Wakai88,Asai84,Sano94,Sato81}) at lower 
$\sqrt{s_{NN}}$ energies ($\sqrt{s_{NN}}$ = 2.7-3.1 GeV).
This work is in a similar spirit but at higher c.m. energy and uses    
different models for (a) $\Lambda$ particles production in the participant
region taking into consideration impact parameter dependence of the
specific collision
and (b) attachment of the $\Lambda$ particle to different 
fragments in the PLF forming what we refer here as composite. 
It is felt that in view of the
future experimental activities 
\cite{HypHI05,HypHI06,Saito07,fukuda07,Gianotti07,Friese07,Toneev07}
the results from alternative models, and at different 
energies, will help to better establish the mechanism for
formation of hypernuclei. 
 
Our calculations are performed in two separate stages. 
For a given impact parameter ($b$) a large number of events 
($10^5-10^7$) are generated with
{\small HIJING/B\=B} model in order to obtain  the average 
number of $\Lambda$ particles 
per event ($<n_{\Lambda}(b)>$) within appropriate 
kinematic cuts in rapidity and momentum. 
Within the model we can also calculate 
the average number of the nucleons in the PLF,  
$<n_{\rm PLF}>$ = $<n_1(b)> + <n_2(b)>$, where 
 $n_1(b)$ stands for neutrons, and $n_2(b)$ for protons.

We can characterize a produced composite (with and 
without strangeness) by three symbols: $a$, $z$ and $h$, where $a$ is
the atomic mass, $z$ is the charge value of the isotope, and $h$ refers
to the number of hyperons embedded into the nucleus ($a$, $z$).
For $h$ = 0 we have normal composites. 
For $h$ = 1 we have a hypernucleus with one
$\Lambda$. It is also possible to include the case 
$h$ = 2 (i.e., $^a_{\Lambda\Lambda}z$).
Similarly it is possible to include other kinds 
of hyperons, i.e., $\Sigma$.
However, this is beyond the scope of the present work.
Calculations using canonical thermodynamic model (CTM)  
\cite{Subal09,Chaudhuri07,Elliott03} are performed in the second stage  
to estimate the average number ($<n_{a,z,1}(b)>$) of  
 hypernuclei ($^{a}_{\Lambda}$z) produced at a given temperature $T$,  
when the PLF has one $\Lambda$, and the average number of nucleons  
$<n_{\rm PLF}>$ =$<n_1(b)> + <n_2(b)>$.

There are two major approximations in our approach:
(i) the time dependence of source function for hyperons and 
fragments has been neglected;
(ii) we neglect also secondary hypernuclear processes.
The transport calculations predict only moderate contributions
\cite{Gaitanos09}
to the total hyper-fragment cross section from indirect coalescence
through the $\pi B$ channel (where B stands for a baryon and $\pi$
for a pion).      
With these assumptions, the formation cross section  
of a hypernucleus in the PLF rest system can be expressed by:

\begin{equation}
\sigma(^{A_{F}}_{\Lambda}Z)= 
\int 2\pi\,b\, db\,<n_{\Lambda}(b)>\,<n_{a,z,1}(b)>
\end{equation}
 
The basic outline and main parameters of {\small HIJING/B\=B} model
are presented in Sec. II. We employ reasonable theoretical evaluations 
for the $\Lambda$ particle production with specific kinematic cuts.
In Sec. III we give details of CTM calculations. 
The results and the discussions are presented in Sec. IV. The final 
conclusions are summarized in Sec. V.

\section{Outline of HIJING/B\=B v2.0 model.}

The {\small HIJING} \cite{miklos_94} and {\small HIJING/B\=B} v1.10 models 
\cite{miklos_99} have been used extensively 
to determine the physical 
properties of the matter produced in rHIC and to study particle production.
String models describe the collisions through the exchange of color or 
momentum between partons in the projectile and target.
As a consequence of these exchanges, these partons become joined by 
colorless objects, which are called string, ropes or flux tubes.
In {\small HIJING} type models \cite{miklos_94} 
the soft beam jet fragmentation is modeled by diquark-quark strings 
with gluon kinks induced by soft gluon radiations. 
Hard collisions are included within perturbative Quantum
Chromodynamics (pQCD) computed parton-parton collisions.
The mini-jets physics 
is embedded via an eikonal multiple collision framework
using pQCD to compute the initial and final state radiation
and hard scattering rates. 
The cross section for hard parton scatterings is enhanced by
a factor K=2 in order to simulate high order corrections. 
The HIJING  model was originally designed for hadron-hadron interactions.
Generalization to hadron-nucleus ($p+A$) and nucleus-nucleus ($A+A$)
collisions is performed \cite{miklos_94} through the Glauber theory.
Besides the Glauber nuclear eikonal extension,
shadowing of nuclear parton distributions
is modeled. In addition,  dynamical energy loss
of the (mini)jets is taken into account through an effective 
dE/dx (approximately 2 Gev/fm per gluon mini-jet).
These models contain a {\em soft} and a {\em hard} component,
which is crucial for their application from FAIR to 
Large Hadron Collider (LHC) energies.

Unlike conventional di-quark fragmentation implemented  
in HIJING model \cite{miklos_94},
a baryon junction allows the di-quark to split with the
three independent flux lines tied together at a junction in 
{\small HIJING/B\=B} v1.10 \cite{miklos_99}.
We introduce \cite{top04_prc} 
a new version (v2.0) of {\small HIJING/B\=B} that differs from 
{\small HIJING/B\=B} v1.10 \cite{miklos_99}  
in its implementation of hypothesized junction anti-junction (J\=J)
loops.
A picture of a junction loop is as follows: a
color flux line splits at some intermediate point
into two flux line at one junction and then the flux line fuse back
into one at a second anti-junction somewhere further 
along the original flux line.  
Before fragmentation, we compute the probability
that a junction loop occurs in the string.
The probability of such loop 
is assumed to increase with the number of binary interactions
that the incident baryon suffers in passing through the 
oncoming nucleus \cite{top04_prc}. This number depends
on the impact parameter and is computed in {\small HIJING}
using eikonal path through a diffuse nuclear density.
Moreover, we add  an enhanced 
intrinsic (anti)di-quark $p_T$ kick in any standard (q-qq) strings that 
should contain one or multiple J\=J loops.
This mechanism correspond to some degree of {\em collectivity} 
for parton interactions.

In string fragmentation phenomenology, it has been proposed
that the strong enhancement of strange particle
observables require strong color field (SCF) effects \cite{miklos_85} .
The particle production is large and dominated by pair production 
and the energy density appears to exceed the one required for 
quark gluon plasma (QGP) formation.
The overall scenario is consistent with particle
production from a strong color field (SCF), formation of a QGP and
subsequent hadronization. The SCF effects are also possible source 
of novel {\em baryon/hyperon physics}.
In our previous works
\cite{top05_prc,top07_prc,armesto_08,top08_prl}, 
we explore with {\small HIJING/B\=B} v2.0 model calculations, 
possible dynamical
effects associated with long-range coherent field (SCF), that
may arise in rHIC.
These analysis point to the {\em need for
collective motion}, together with {\em high initial-state energy densities}.

For a uniform chromoelectric flux tube with field ({\it E}) 
the pair production rate \cite{schwinger}  
per unit volume for a heavy quark is given by:
\begin{equation}
\Gamma =\frac{\kappa^2}{4 \pi^3} 
{\text {exp}}\left(-\frac{\pi\,m_{Q}^2}{\kappa}\right)
\end{equation}
where for strange quark $Q=s$, the current quark mass 
is in the range of $m_{\rm s}=80-190$ MeV and 
the constituent quark mass $M_{\rm s} = 350$ MeV 
(for detailed discussion see Ref.~\cite{top07_prc}). 
Note that $\kappa=|eE|_{eff}= \sqrt{C_2(A)/C_2(F)}\, \kappa_0$ 
is the effective string tension
in terms of the vacuum string tension $\kappa_0 \approx 1 $ GeV/fm
and $C_2(A)$, $C_2(F)$ are the second order Casimir operators 
(see Ref. \cite{miklos_85}).
A measurable rate for spontaneous pair production requires 
``{\em strong chromo electric fields}'', such 
that $\kappa/m_{\rm Q}^2\,\,>$ 1 {\em at least some of the time}.
On the average, longitudinal electric field ``string'' models 
predict for heavier flavor a very suppressed production rate per unit
volume $\gamma_{Q}$ via the well known Schwinger formula \cite{schwinger}, 
since
\begin{equation}
\gamma_{Q\bar{Q}} = \frac{\Gamma_{Q\bar{Q}}}{\Gamma_{q\bar{q}}} =
{\text {exp}} \left(-\frac{\pi(m_{Q}^2-m_q^2)}{\kappa_0} \right)
\ll 1
\end{equation}
for $Q=s$ and $q=u,d$.
For a color rope on the other hand,
if the {\em average} string tension value ($<\kappa>$) 
increases from 1.0 GeV/fm to 2.0-3.0 GeV/fm,
the rate $\Gamma$ for strangeness pairs 
to tunnel through the longitudinal field increases
(see Refs. \cite{top05_prc,top07_prc}).

We take into account SCF in our model by an 
{\em in medium effective string tension} 
$\kappa > \kappa_0$, which lead to new values for the suppression factors, 
as well as the new effective intrinsic transverse momentum $k_T$. 
This includes: 
i) the ratio of production rates of  
di-quark to quark pairs (di-quark suppression factor),  
$\gamma_{{\rm qq}} = P({\rm qq}\overline{{\rm qq}})/P(q\bar{q})$,
ii) the ratio of production rates of strange 
to non-strange quark pairs (strangeness suppression factor), 
$\gamma_{s} = P(s\bar{s})/P(q\bar{q})$,
iii) the extra suppression associated with a diquark containing a
strange quark compared to
the normal suppression of strange quark ($\gamma_s$),
$\gamma_{us} = (P({\rm us}\overline{{\rm us}})/P({\rm ud}\overline{{\rm ud}}))/(\gamma_s)$,
iv) the suppression of spin 1 diquarks relative to spin 0 ones
(apart from the factor of 3 enhancement of the former based on
counting the number of spin states), $\gamma_{10}$, and 
v) the (anti)quark ($\sigma_{q}'' = \sqrt{\kappa/\kappa_0} \cdot \sigma_{q}$)
and  (anti)di-quark ($\sigma_{\rm {qq}}'' = \sqrt{\kappa/\kappa_0}
\cdot {\it f} \cdot \sigma_{{\rm qq}}$) Gaussian  width.

In this paper we extend our study in the framework of  
{\small HIJING/B\=B v2.0} model \cite{top07_prc}
to strange particle production at the FAIR and MPD/NICA energy range
($\sqrt{s_{NN}}$ = 2.7-9.4 GeV).
The experimental situation has so far been rather poor
for measurements of strange particles at forward rapidities,
due to limited acceptances of the detectors. The very forward 
rapidity (closer to projectile rapidity) are dealt with by models only. 
Using {\small HIJING/B\=B v2.0} model we analyze the production 
of the average value per event  $<n_{\Lambda}(b)>$
with specific kinematic cuts in rapidity 
$(y_{proj} - 0.05) \leq \delta y \leq (y_{proj} + 0.05)$,
(where $y_{proj}$ stands for projectile rapidity)
and total momentum $p_{\rm tot} < 250$ MeV/c
in the PLF rest system, which can leads to the formation of 
hypernucleus in rHIC.
The main parameters used in the calculations presented here are given
in Table \ref{tab:tab1}.

\begin{table}
\caption{Main parameters used in symmetric $A + A$ collisions. 
The parameters are 
defined in the text. Set 1 correspond to calculations without J\=J loops
and SCF effects. 
Set 2 and 3 include both effects and correspond to different mean values
of the string tension.}
\begin{ruledtabular}
\begin{tabular}{ccccccccc}    
A + A & $\kappa$ & $ \gamma_{{\rm qq}}\,\,$ & 
$\gamma_{s}\,\,$ & $\gamma_{{\rm us}}\,\,$ & $\gamma_{10}\,\,$ & 
 $\sigma_{q}\,\,$ & $f\,\,$ & \\
    & (GeV/fm) &   &   &   &   & (GeV/c) &  & \\
\hline
Set 1 & $\kappa_0$ = 1.0 & $\,\,0.02\,\,$ & $\,\,0.30\,\,$ &
$\,\,0.40\,\,$ & $\,\,0.05\,\,$ & $\,\,0.350\,\,$ & $\,\,1\,\,$ & \\
Set 2 & $\kappa_2$ = 2.0 & $\,\,0.14\,\,$ & $\,\,0.55\,\,$ & $\,\,0.63\,\,$ & $\,\,0.12\,\,$ & $\,\,0.495\,\,$ & $\,\,3\,\,$ & \\
Set 3 & $\kappa_3$ = 3.0 & $\,\,0.27\,\,$ & $\,\,0.67\,\,$ & $\,\,0.74\,\,$ & $\,\,0.18\,\,$ & $\,\,0.606\,\,$ & $\,\,3\,\,$ & \\
\end{tabular}
\end{ruledtabular}
\label{tab:tab1}
\end{table}

In these calculations we do not consider nuclear effects
such as shadowing and quenching \cite{miklos_94}, \cite{top04_prc},
which are specific only for very high energy
interactions ($\sqrt{s_{NN}} \geq 20$ GeV).
Within our model we do not include 
rescattering processes which Relativistic Quantum Molecular Dynamics 
(RQMD) model calculations \cite{lambda_ags01} 
show to be negligibles at the forward rapidity. 
The physics embedded in our model include 
a ``hard''(pQCD) and an underlying ``soft'' part, therefore we can not
apply our phenomenology for energies $\sqrt{s_{NN}} \leq $ 3.0 GeV.
The detailed analysis for (multi)strangeness 
production from FAIR to SPS and RHIC energies within 
{\small HIJING/B\=B v2.0} model will be presented elsewhere
\cite{top_2010}.

\begin{figure}[t]
\includegraphics[width=3.5in,height=1.75in,clip,angle=0]{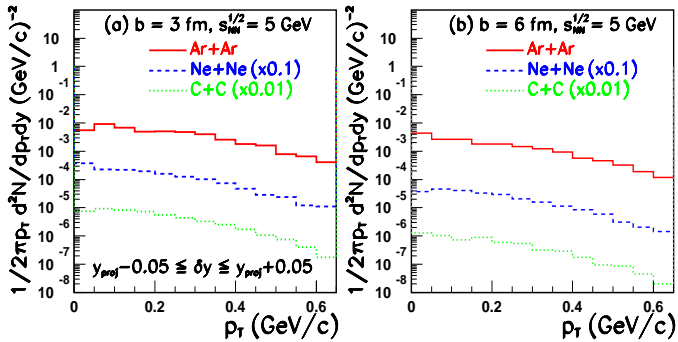}
\caption{(color online). The {\small HIJING/B\=B v2.0} model predictions for  
$p_T$ spectra of $\Lambda$s particle within rapidity cut
$\delta y$ at $\sqrt{s_{NN}}$ = 5 GeV.
For clarity the values for Ne + Ne and C + C are multiplied by
0.1 and 0.01 respectively.}
\label{fig:fig1}
\end{figure}

For a given impact parameter ($b$) we 
calculate with {\small HIJING/B\=B v2.0} model
the average number per event $<n_{\Lambda}(b)>$, in appropriate
kinematic phase space, and the average number of nucleons 
in the PLF , $<n_{PLF}(b)>$.
The predictions for transverse momentum ($p_T$)
spectra of $\Lambda$ particles
with rapidity cut $\delta y$ for symmetric colliding systems 
$^{12}$C + $^{12}$C, $^{20}$Ne + $^{20}$Ne, 
and $^{40}$Ar + $^{40}$Ar at $\sqrt{s_{NN}}$ = 5 GeV
are presented in Fig.~\ref{fig:fig1} for central 
($b$ = 3 fm, left panel) and peripheral ($b$ = 6 fm, right panel).
The values of the parameters 
used to obtain the results discussed here, 
are Set 2 from Table~\ref{tab:tab1}.

The average value per event
$<n_{\Lambda}(b)>$ within rapidity cut $\delta y$ and 
with total momentum ($p_{\rm tot} < 250$ MeV/c) 
in the projectile rest system,  
are given in Table \ref{tab:tab2} for the symmetric colliding system 
$^{12}$C + $^{12}$C, $^{20}$Ne + $^{20}$Ne
and $^{40}$Ar + $^{40}$Ar. 
These numbers are the main input in Eq. 1 used to
calculate inclusive cross sections for the formation of hypernuclei
in nucleus-nucleus collisions in the PLF rest frame.

\begin{table}
\caption{The dependence on the impact parameter ($b$) of the 
mean number of $\Lambda$'s per event ($<n_{\Lambda}(b)>$)  
obtained within kinematic cuts (see Sec.~II) 
in the PLF rest frame for symmetric colliding systems:
$^{12}$C + $^{12}$C, $^{20}$Ne + $^{20}$Ne
and $^{40}$Ar + $^{40}$Ar.}
\begin{ruledtabular}
\begin{tabular}{ccccc}    
       & C + C & Ne + Ne & Ar + Ar  & \\

 $\sqrt{s_{NN}}$ &  3.7 GeV&  5.0 GeV& 5.0 GeV & \\
  $b$ (fm) & $<n_{\Lambda}^{\rm C}(b)>\times 10^4$   & 
  $<n_{\Lambda}^{\rm Ne}(b)>\times 10^4$     &
  $<n_{\Lambda}^{\rm Ar}(b)>\times 10^4$   & \\
\hline
 0.0 & 4.61 & 6.70 & 16.4 & \\
 1.0 & 2.80 & 6.00    & 16.20     & \\
 2.0 & 3.01 & 5.60    & 18.90    & \\
 3.0 & 1.44 & 5.20    & 15.80    & \\
 4.0 & 0.38 & 3.40    & 11.80    & \\
 5.0 & 0.42 & 2.00    & 9.92    & \\
 6.0 & 0.24 & 0.77    & 4.26    & \\
 7.0 & 0.063 &0.42    & 2.74    & \\
 8.0 & 0.088 &0.25    & 1.62    & \\
 9.0 &      & 0.095   & 0.49    & \\
10.0 &      &         & 0.23    &\\
11.0 &      &         & 0.13    &\\
\end{tabular}
\end{ruledtabular}
\label{tab:tab2}
 \end{table}

Our estimate for total number of events with one 
$\Lambda$ particle produced in the projectile rapidity region in 
C + C collisions at $\sqrt{s_{NN}}$ = 3.7 GeV is $13.05 \,\times \,10^{-4}$.
A comparable value ($26.6 \,\times\, 10^{-4}$) 
at lower energy (2A GeV in laboratory system),  
is obtained \cite{HypHI06} 
with ultrarelativistic quantum molecular dynamics (UrQMD) 
calculations by assuming a coalescence of the produced 
$\Lambda$ hyperons with spectators.
However, their rapidity cut is much larger 
$(y_{proj} - 0.4) \leq \delta y_{\rm QMD} \leq (y_{proj} + 0.4)$, 
than that used in our calculations.
Therefore, we investigate within {\small HIJING/B\=B v2.0} model
the sensitivity of our predictions to different rapidity cuts $\delta y$.
An increase of $\delta y$ from 0.05 to $\delta y$ = 0.5
results in an increase of a factor 6-8 for the 
average value $<n_{\Lambda}(b)>$.   
The sensitivity to the mean values of the in medium string tension
($\kappa$) is less important. Using Set 3  
from Table \ref{tab:tab1} which correspond to 
a mean value $\kappa = 3$ GeV/fm, 
result in a modest increase (less that $3-5\, \%$) of $<n_{\Lambda}(b)>$,
for Ne + Ne central collisions ($b$ = 3 fm) at $\sqrt{s_{NN}}$ = 5 GeV.

Finally, for the symmetric colliding system 
$^{12}$C+$^{12}$C, we investigate also 
the energy dependence of particle production 
for central collisions ($b$ = 3 fm). 
When $\sqrt{s_{NN}}$ increase from 3.7 GeV to 5 GeV, the number of 
$\Lambda$ particles ($<n_{\Lambda}(b)>$) within kinematic cuts discussed above,
has a modest increase of only 32 $\%$. 
The model predict a more dramatic increase 
of $<n_{\Lambda}(b)>$, i.e., a factor of 2.75 when   
$\sqrt{s_{NN}}$ increase from 5 GeV to 10 GeV. 
Since the energy range of the onset of deconfinement is expected to be
at $\sqrt{s_{NN}}$ = 6-8 GeV, the above energy range (5-10 GeV) 
should be carefully investigated by the future planned experiments.

\section{Model for hypernucleus formation in the PLF}

The $\Lambda$ particles with appropriate rapidity find themself in the PLF's.
They thermalises along with the nucleons.  In an event the $\Lambda$
hyperon can
remain unattached to nucleons or can combine with some nucleons to form
a hypernucleus.  There will be also ``normal'' composites, those without
strangeness.  Assuming equilibration we can compute the average
numbers of both normal composites and hypernuclei.  
Past experience has shown that
the temperature in the PLF is expected to be in the range of 5 to 8 MeV 
\cite{Chaudhuri07,Elliott03}.

The Canonical Thermal Model (CTM) for two kinds of 
particles (neutron and proton) has had long usage  
\cite{Das1}. The extension to three
kinds of particles (neutron, proton and $\Lambda$) was discussed by us 
in Ref. \cite{Subal09}.
In this analysis we just give the details necessaries to follow 
the calculations performed here.
In CTM we need to calculate the partition function $Q$:
\begin{equation}
Q_{A_{F},Z,H}=\sum\prod\frac{(\omega_{a,z,h})^{n_{a,z,h}}}{n_{a,z,h}!}
\end{equation} 
where $A_{F}$ is the number of nucleons in the PLF plus one 
(the $\Lambda$ particle); $Z$ is the number
of protons in the PLF and $H$ = 1 (only one $\Lambda$ with appropriate 
kinematic cuts discussed in Sec.~II 
entered the PLF); $h$ = 0 (normal composites) or $h$ = 1, a hypernucleus;
$\omega_{a,z,h}$ is the partition function of one composite with 
quantum numbers $a,z$ and $h$; $n_{a,z,h}$ is the number of this composite 
in a given channel.
The sum in the above equation includes an enormous number of channels
which satisfy conservation of quantum numbers:
\begin{eqnarray}
\sum an_{a,z,h} &=& A_{F}  \nonumber \\
\sum zn_{a,z,h} &=& Z  \nonumber  \\
\sum hn_{a,z,h} &=& H
\end{eqnarray}
It is shown in Ref. \cite{Subal09} that the average occupation of $(a,z,h)$
is given by:
\begin{equation}
<n_{a,z,h}>=\frac{1}{Q_{A_{F},Z,H}}\omega_{a,z,h}Q_{A_{F}-a,Z-z,H-h}
\end{equation}
The partition functions $Q$ can be calculated using recursion relations
discussed in Ref. \cite{Subal09}.

The one particle partition function $\omega_{a,z,h}$ has two parts:
\begin{equation}
\omega_{a,z,h}=z_{kin}(a,z,h)\,z_{int}(a,z,h)
\end{equation}
The kinetic part is given by
\begin{equation}
z_{kin}(a,z,h)=\frac{V}{h^3}(2\pi MT)^{3/2}
\end{equation}
where $V$ is the freeze-out volume in which thermodynamics is assumed. 
For an atomic mass $a$ we take V as 3 times 
the normal nuclear volume 
( $V_0=a/\rho_0$, with $\rho_0 \approx $ 0.15 fm$^{-3}$), but then   
this volume is reduced for Van der Waals excluded volume
($V = 2 V_{0}$). 
The mass of the composite $M$ is:
$M=(a-h)m_n + hm_{\Lambda}$, where $m_n$ (938 MeV) and $m_{\Lambda}$
(1116 MeV) stand for the nucleon mass and $\Lambda$ mass respectively. 

The internal partition function,
$z_{int}(a,z,h)$ could be written as: 
\begin{equation}
z_{int}(a,z,h)=\sum_i(2s_i+1)exp(-\beta e_i)
\end{equation} 
where $e_i$ are energy eigenvalues of the composite $(a,z,h)$.  However,
considerable caution is needed in taking the sum over the energy states $i$.
Because of Levinson's theorem the sum needs to be modified
and cut off (see the discussion on Ref. \cite{Koonin}).
In the previous work \cite{Subal09} the interest was 
to investigate gross features and a relative production only was 
discussed. Therefore, 
a generic formula for ground state binding energy and excited states was
used for most nuclei.  

In this analysis we are attempting a more quantitative estimate 
for production cross sections of hyperfragments $^{A_{F}}_{\Lambda}Z$ 
for $Z$ = 1 to 4.  
The calculated cross sections differ significantly if we use the generic
formula from Ref.~\cite{Subal09}, 
or experimental energies for ground and excited states.
In the calculations performed here we have
taken experimental values of ground state and excited state energies for 
low mass hypernuclei ($a\,<\,11$).  
We include only particle stable excited states.
The binding energies of ground states
for composites with strangeness $(a,z,1)$ are
estimated from tabulated values of $B_{\Lambda}$ 
and binding energy of normal composites $(a-1,z,0)$.    
Data for $a\,<\,11$ are taken
from Refs.~\cite{Oshimata,Bando90} and references therein. The details
are:
 
$^3_{\Lambda}$H: only 1 state with spin 1/2.

$^4_{\Lambda}$H: ground state with spin 0 and one excited state with
spin 1 at 1.04 MeV excitation.

$^4_{\Lambda}$He: ground state with spin 0 and one excited state with
spin 1 at 1.15 MeV excitation.

$^5_{\Lambda}$He: only ground state with spin 1/2.

$^6_{\Lambda}$He: ground state with spin 1.

$^6_{\Lambda}$Li: ground state with spin 1.

$^7_{\Lambda}$He: ground state with spin 1/2.

$^7_{\Lambda}$Li: ground state with spin 1/2, excited state spin 3/2
at 0.69 MeV, excited state spin 5/2 at 2.05 MeV, excited state spin 7/2 at
2.521 MeV, excited state spin 1/2 at 3.56 MeV.

$^7_{\Lambda}$Be: ground state with spin 1/2.

$^8_{\Lambda}$Li: ground state with spin 1, one excited
state with spin 1 at 1.22 MeV excitation

$^8_{\Lambda}$Be: ground state with spin 1, one excited state with
spin 1 at 1.22 MeV excitation.

$^9_{\Lambda}$Li: ground state with spin 3/2.

$^9_{\Lambda}$Be: ground state with spin 1/2 and excited states with
spins 5/2 and 3/2 grouped at 2.93 MeV.

$^{10}_{\Lambda}$B: ground state with spin 1, excited states with spin 2 at
0.22 MeV, another with spin 2 at 2.47 MeV and a spin 3 at 2.70 MeV.

$^{10}_{\Lambda}$Be: experimental binding is used but excitation 
energies and spins are taken to be the same as for $^{10}_{\Lambda}$B
by appealing to isospin symmetry.

For heavier hypernuclei ($a > 10$), a liquid-drop formula is used for 
ground state energy.  This formula is taken from Ref.~\cite{Botvina07}.
All energies are in MeV.
\begin{widetext}
\begin{equation}
e_{a,z,h}=-16a+\sigma(T)a^{2/3}+0.72z^2/(a^{1/3})+25(a-h-2z)^2/(a-h)
-10.68h+21.27h/(a^{1/3})
\end{equation}
\end{widetext}
where $\sigma(T)$ is the surface tension which is a function dependent 
on temperature:
\begin{equation}
\sigma(T)=18\left[\frac{T_c^2-T^2}{T_c^2+T^2}\right]^{5/4} 
\end{equation}
A comparative study
of the above binding energy formula can be found in Ref.~\cite{Botvina07}.
This formula also defines the drip lines.  We include all nuclei
within drip lines in constructing the partition function.

In order to calculate $z_{int}(a,z,h)$ we use 
the liquid-drop formula for $e_{a,z,h}$ and include also  
the contribution coming from the excited states.
This results in a
multiplicative factor exp (r(T)Ta/$\epsilon_0$), where 
$r(T)=12/(12+T)$ is a correction term.
For a detailed discussion of parameters used in these formulae  
see Ref~\cite{Bondorf}.  
As temperature $T$ increases,  
this correction slows down the increase of partition 
function $z_{int}(a,z,h)$  
due to excited states. Similar correction has been used before
\cite{Bhattacharyya,Koonin}, although this is not important for the 
temperature range considered here.
We note, that our calculations take into consideration  
the effects of the long-range Coulomb force
in the Wigner-Seitz approximation \cite{Bondorf}.

Apart from hypernuclei, we need also to specify the 
partition function $\omega$ for
normal composites ($h$ = 0). For $^1$H, $^2$H, $^3$H, $^3$He,
$^4$He, $^5$He, $^5$Li, $^6$He, $^6$Li and $^6$Be we use experimental
ground state energy and no excited states. For atomic mass 
$a\,>\,6$ we use the generic formula from Eq. 10 with $h$ = 0 
and consider also the 
contributions from the excited states as described above.

The temperature ($T$) in the PLF is also an important parameter
in our calculations and it is expected to have no dependence on the  
beam energy value (in the range of 3-10 GeV).
Many data can be used to estimate the temperature and  
the range $T$ = 5 MeV to $T$ = 8 MeV is a 
reasonable one in describing these data 
\cite{Chaudhuri07}, \cite{Das1}. 
Our results show, that predicted cross sections 
can change considerably within this range of the 
input temperatures, since at lower temperature
heavier hypernuclei are favored at the expense of lighter ones.  
The average occupation number $<n_{a,1}>$ = $\sum_z$$ ^a_{\Lambda}z$
will first drop with $a$, go through a minimum and rise again
(see Fig.~1 and Fig.~2 from Ref.~\cite{Subal09}).
This is the well-known {\em U} shape. As the temperature changes,
the shape changes very quickly to a different one, where the
occupation falls monotonically with increasing $a$.
The temperature domain of 5 MeV to 8 MeV is precisely the range
where this happens.
As a results occupations of the heavier constituents in the 
PLF can change dramatically. 
These suggest that the relative populations of lighter hypernuclei 
could be a good ``measure'' of the temperature in the PLF.
Such effects are also seen in our results discussed in the
Sec. IV (see Table~\ref{tab:tab3}), but the situation there 
is more complex, because of 
impact parameter dependence and the expected fluctuations on
the nucleon numbers in the PLF.

We also note, that the PLF system is finite, and in this temperature range 
grand canonical model results differ 
drastically from those obtained within CTM model,
which include additional constrains such as particles number
conservation.
The detailed comparison of the results obtained using both 
models can be found in Ref.~\cite{Das1}.

\section{Results and Discussions}

This work presents the inclusive production cross sections
of different types of hypernuclei for the symmetric colliding systems 
$^{12}$C + $^{12}$C at $\sqrt{s_{NN}}$ = 3.7 GeV, $^{20}$Ne + $^{20}$Ne
and $^{40}$Ar + $^{40}$Ar at $\sqrt{s_{NN}}$ = 5 GeV.  
We assume that $\Lambda$ particles with a rapidity cut 
$\delta y $ and total momenta less than 250 MeV/c in the rest frame of the
projectile will thermalises with the nucleons in the PLF and will produce
hypernuclei. In all three cases studied here 
the following characteristics are common.
Integration over impact parameter is important. The relevant quantity is
the product $b<n_{\Lambda}(b)>$ in Eq. 1. 
The peak of this function is not too sharp. For small
impact parameters ($b < 4$ fm) the number of nucleons 
in the PLF is small and thus
only light hypernuclei can be formed when the hyperon $\Lambda$ is captured.
For large impact parameters ($b \geq 4$ fm) 
the number of nucleons in the PLF is greater than in the above case,
and both light and heavy hypernuclei are formed.

Previous theoretical studies through coalescence model
\cite{Wakai88,Asai84,Sano94,Sato81} predicted the cross sections 
of the order of few microbarn ($\mu b$). 
The inclusive cross sections obtained in our model 
for different types of hypernuclei (as
indicated) for the light colliding system $^{12}$C + $^{12}$C at 
$\sqrt{s_{NN}}$ = 3.7 GeV are given 
in Table \ref{tab:tab3} for temperature $T = 5$ MeV (second column) and 
$T = 8$ MeV (third column) in comparison with coalescence model results
(at $ \sqrt{s_{NN}}$ $\approx $ 2.7 GeV) \cite{Wakai88,Asai84,Sano94,Sato81}. 
One knows that the total cross section 
of nucleon-nucleon ($N-N$) scattering
becomes nearly constant at aproximately 40 mb for c.m. energies
in the range 2.5-10 GeV, and in this energy region only 
$1\, \%$ of the ($N-N$) scattering
produces a baryon of strangeness 1. Also we note that there is an energy 
threshold of $\approx$ 1.6 GeV for $\Lambda$ production 
in an elementary process of $NN\,\rightarrow \,\Lambda\,K\,N$.
Therefore, we could consider this comparison as appropriate, since in
the region $ \sqrt{s_{NN}}$ = 2-4 GeV, the $\Lambda$ hyperons
production is expected to have a modest increase.
Moreover, for Ne + Ne collisions, the calculations 
with coalescence model \cite{Wakai88} show that the hypernucleus 
formation cross sections have also a modest increase (by 30-50 $\%$), 
when $ \sqrt{s_{NN}}$ increase
from 2.04 GeV to 3.1 GeV (see Fig.~6 in Ref.~\cite{Wakai88}).

\begin{table}
\caption{Inclusive production cross sections (all values are in $\mu b$) 
for different types of 
hypernuclei for the colliding system $^{12}$C + $^{12}$C
at $\sqrt{s_{NN}}$ = 3.7 GeV. Our results (column 2 and 3) are compared with
previous predictions obtained within coalescence model 
\cite{Bando90,Bando2,Sano94} at $\sqrt{s_{NN}}$ $\approx$ 2.7 GeV.}
\begin{ruledtabular}
\begin{tabular}{ccccc}    
 Model  & Coalescence & CTM       & CTM    & \\
 Hypernuclei &        & (T = 5 MeV)    &  (T = 8 MeV)      &\\
\hline
 $^3_{\Lambda}$H & 0.22 & 0.89  &  3.25 &\\
 $^4_{\Lambda}$H &  0.39 & 0.32 & 0.71 &\\
 $^4_{\Lambda}$He & 0.39 &  0.34 &  0.77 & \\
$^5_{\Lambda}$He &  2.58 &   3.87 &  1.46 & \\
  $^6_{\Lambda}$He &  0.32 &  0.50 & 0.17 & \\
$^7_{\Lambda}$He &   0.09   &   0.0009 & 0.004& \\ 
 $^6_{\Lambda}$Li & 0.30  &  0.56  &   0.18& \\
 $^7_{\Lambda}$Li &  0.24 & 26.88  &  0.85& \\
 $^8_{\Lambda}$Li &  0.18 & 0.17  &  0.13& \\ 
 $^9_{\Lambda}$Li &  0.05 & 0.00008 & 0.0004& \\
$^7_{\Lambda}$Be &   0.07 &  0.001  &  0.005& \\
$^8_{\Lambda}$Be &  0.15  &  0.18   &  0.13 & \\
$^9_{\Lambda}$Be & 2.48  & 22.3   &  2.26 & \\
$^{10}_{\Lambda}$Be & 0.33 &  0.02 &  0.018\\
\end{tabular}
\end{ruledtabular}
\label{tab:tab3}
 \end{table}

Our results are different from those reported 
within the coalescence model \cite{Wakai88,Asai84,Sano94,Sato81} which was used
often to estimate cross sections for hypernucleus formation in the PLF.
Only for few isotopes (e.g.,$^4_{\Lambda}$He, 
$^8_{\Lambda}$Li,$^8_{\Lambda}$Be) we obtain inclusive cross sections 
slightly different in comparison with those predicted 
by the coalescence model. 
It is worthwhile to highlight the important differences.

To understand the basic idea of the coalescence model,
let us begin by first considering a PLF without a hyperon.  It is  
excited (usually parametrized by
a temperature) and breaks up (as is well established experimentally)
into many fragments.  The distribution of these fragments ($F$) in the PLF can
be denoted by $\frac{d^3N^F}{d^3p_F}(\vec p_F)$.  
Usually this distribution
is not calculated in the coalescence model for hypernucleus formation 
although in some version of the coalescence this could be attempted.
In principle, this distribution should be taken from experiment.
In the present problem, we have, in addition, 
a $\Lambda$ particle with a momentum distribution. 
If the velocity of the fragment 
$\vec p_F/m_F$ and the velocity of the $\Lambda$ particle,
$\vec p_{\Lambda}/m_{\Lambda}$ match they can coalesce into a hypernucleus.
Thus the cross-section is given by the product of two distributions and
an overlap factor which gives the probability that the $\Lambda$ and $F$ 
becomes the hypernucleus $_{\Lambda}F$ 
(see Eq. 15 in Ref. \cite{Wakai88} and Eq. 11 in 
Ref. \cite{Asai84}).
Note that the $\Lambda$ plays a very passive role here.
Practitioners of the coalescence model suggested 
that $\frac{d^3N^F}{d^3p_F}$ has to be taken from experiment.  
In the past applications of the coalescence model, experimental values of
$\frac{d^3\sigma^F}{d^3p_F}$ were not used as not all of those needed
for predicting hypernucleus production were experimentally available.
A theoretical model parametrisation for $\frac{d^3\sigma^F}{d^3p_F}$ had
to be employed. These parameterizations give (wrongly) finite 
values for inclusive cross-sections of $^5$He and $^8$Be nuclei 
and hence finite values for inclusive cross-sections formation for both 
$^6_{\Lambda}$He and $^9_{\Lambda}$Be.  But this
is fortuitous, since $^5$He and $^8$Be nuclei are
not bound systems by themselves \cite{Carsten98,Yamada09}.
However, calculations based on a generalized mass formula for 
non-strange and hypernuclei predict the existence of several 
bound hypernuclei (e.g.,$^6_{\Lambda}$He and $^9_{\Lambda}$Be) 
whose normal cores are unbound \cite{samanta2006,samanta2008}.

\begin{figure}[h]
\includegraphics[width=3.5in,height=1.75in,clip,angle=0]{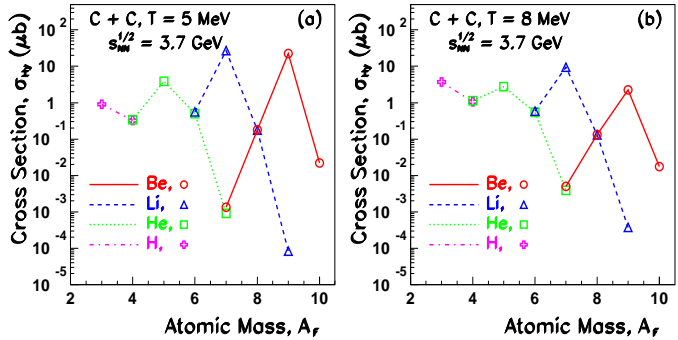}
\caption{(color online). The hypernucleus formation cross 
sections ($\sigma_{Hy}$) in 
$^{12}$C + $^{12}$C collisions at $\sqrt{s_{NN}}$ = 3.7 GeV as a function of 
nuclear fragment mass number $A_F$. The results (open symbols) are shown for 
H, He, Li and Be isotopes and are obtained for  
T = 5 MeV (left panel) and T = 8 MeV (right panel).
The lines are only to guide the eyes.}
\label{fig:fig2}
\end{figure}

These remarks lead to some features which are experimentally verifiable.
For example, correct application of coalescence principles  
should gives zero cross section 
for $^6_{\Lambda}$He and $^9_{\Lambda}$Be isotopes, 
since these require separately a $^5$He nucleus
and a $\Lambda$ or a $^8$Be nucleus and a $\Lambda$. 
The extra role the $\Lambda$ play as ``glue'' 
ensuring stability, can not be incorporated in the coalescence model.
In contrast in CTM model the $\Lambda$ plays a more fundamental role
than in coalescence.  It participates in the thermalization.
The CTM model uses directly the property of $_{\Lambda}$F rather 
than that of $F$ and $\Lambda$ separately.  
Thus both $^6_{\Lambda}$He and 
$^9_{\Lambda}$Be are expected to have non-zero production cross sections.
In fact, we predict a large cross section for $^9_{\Lambda}$Be.
The values obtained within our phenomenology 
are also shown in Fig.~\ref{fig:fig2}(a) (T = 5 MeV) and
in Fig. \ref{fig:fig2}(b) (T = 8 MeV).
Note that the high value reported here for isotope $^7_{\Lambda}$Li
(on dashed lines) could be explained by many low-lying bound states 
of this isotope, which 
all contribute in the calculation of the partition function, $z_{int}$.

\begin{figure}[h]
\includegraphics[width=3.5in,height=1.75in,clip,angle=0]{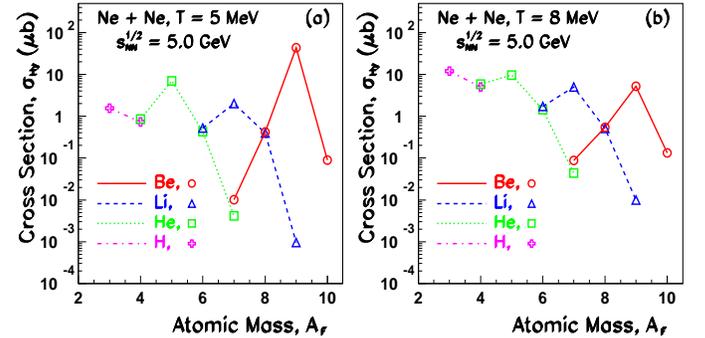}
\caption{(color online). The hypernucleus formation cross sections in 
$^{20}$Ne + $^{20}$Ne collisions at $\sqrt{s_{NN}}$ = 5 GeV.
The open symbols and the lines have the same meaning as in
Fig. \ref{fig:fig2}.}
\label{fig:fig3}
\end{figure}

\begin{figure}[h]
\includegraphics[width=3.5in,height=1.75in,clip,angle=0]{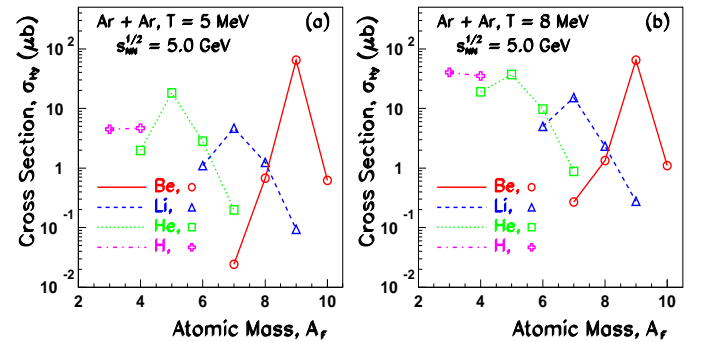}
\caption{(color online). The hypernucleus formation cross sections in 
$^{40}$Ar + $^{40}$Ar collisions at $\sqrt{s_{NN}}$ = 5 GeV.
The open symbols and the lines have the same meaning as in
Fig. \ref{fig:fig2}.}
\label{fig:fig4}
\end{figure}

In collisions between heavy nuclei we expect an increase 
of the production rate of strangeness 
(see Table \ref{tab:tab2} in Sec.~II) and 
of secondary interactions.
In this case the temporal distribution of the fragment and 
$\Lambda$ particle could play an important role and 
the reduction factors of about 0.5 was estimated with coalescence 
model \cite{Sano88} for $^{20}$Ne + $^{20}$Ne at $\sqrt{s_{NN}}$ = 3.1 GeV.
Our results have been obtained by neglecting 
the above time dependence for $^{20}$Ne + $^{20}$Ne 
and $^{40}$Ar + $^{40}$Ar collisions at 
$\sqrt{s_{NN}}$ = 5 GeV. 
Therefore, these values should be considered only as    
upper bounds for formation cross sections of hypernuclei. 
The predictions obtained within our hybrid model 
are shown through graphs in Fig.~\ref{fig:fig3} and Fig.~\ref{fig:fig4}
(left panel, $T$ = 5 MeV) and (right panel, $T$ = 8 MeV).

 The Ar + Ar system is twice as big as Ne + Ne 
(the total number of 80 nucleons versus 40).   
If we consider the results obtained for Li isotopes with 
$T$ = 5 MeV (left panel of Fig.~\ref{fig:fig3} and  Fig.~\ref{fig:fig4}), 
the graphs show that the hypernuclei formation cross sections
are in Ar + Ar system approximately twice than those 
obtained in Ne + Ne system.
In contrast, the production of $^9_{\Lambda}$Be is about the same in both. 
One common feature of all three graphs 
(see Fig.~\ref{fig:fig2}, Fig.~\ref{fig:fig3}, 
 Fig.~\ref{fig:fig4}) is that, the cross section for formation of 
 $^9_{\Lambda}$Be is large and is an order of magnitude higher than
 those predicted within coalescence model and this prediction 
 should be easily verified by future experiments 
 planed at FAIR and MPD/NICA.

Finally, we note that the previous experimental data for the 
production cross sections 
of $^4_{\Lambda}$H and $^3_{\Lambda}$H hypernuclei 
\cite{avramenko92,averyanov08} have been obtained with very light
beams (He, Li) at 3.7A GeV (laboratory system) on carbon fixed target.
We can not study using our approach   
these light colliding systems, since they  
are out of the limit of applicability within our phenomenology. 
To compare our predictions for 
formation cross sections of hypernuclei, data obtained with
heavier beams (targets) and at higher energies are needed.

\section{Conclusions}

We have estimated the 
inclusive production cross-sections of selected light hypernuclei
in the symmetric colliding systems (C + C, Ne + Ne, Ar + Ar) using a hybrid
type model. 
The calculations are performed considering a product of two factors 
in impact parameter space (see Eq. 1).  
The first factor the average value per event, $<n_{\Lambda}(b)>$, estimates 
the production of $\Lambda$ particles in the participant region within a 
limiting phase space. Imposing reasonable kinematic cuts these 
particles could be captured in PLF. The second factor, $<n_{a,z,1}(b)>$, 
describes how these hyperons distribute them-self among the 
different hypernuclei within CTM approach.
The absolute values of cross-sections are basically dependent upon the
first part; the relative values are fixed by the canonical thermodynamic 
model (CTM). When data will become available, these can 
separately test the validity and accuracy of each part.

In comparison with previous predictions within coalescence models,
our results show a striking difference especially for  
$^6_{\Lambda}$He and $^9_{\Lambda}$Be hypernuclei.
A correct application of the coalescence principles should gives 
zero production cross sections for these isotopes. 
In contrast, our hybrid model predict a formation cross sections 
which are significantly larger than 
those reported with coalescence calculations.

Moreover, our approach for the production of hypernuclei in rHIC
also applies with appropriate changes in rapidity and momentum cuts,
to asymmetric colliding systems.
However, to better test our predictions, experimental data obtained 
with heavier beams (target) and in the energy region of interest
$\sqrt{s_{NN}}$ = 5-10 GeV are required.

\section{Acknowledgment}

\vskip 0.2cm

We thank J. Barrette and M. Gyulassy for 
many fruitful and stimulating discussions.
One of us (VTP) thanks M. Gyulassy for a pleasant hospitality
at Columbia University, New York, where part of these calculations
have been performed. 
This work was supported by the Natural Sciences and Engineering 
Research Council of Canada.

\end{document}